\documentclass{article}
\usepackage{amsmath}
\usepackage{amsfonts}
\usepackage{amssymb}
\usepackage{amsthm}
\usepackage{times}
\usepackage{cancel}
\usepackage{dsfont}
\usepackage{graphics}
\usepackage[T1]{fontenc}

\begin{document}
\date{}
\title{\flushleft\textbf{How Can It Be Like That?}\\\bigskip \textbf{JEFFREY BUB\thanks{Department of Philosophy,
Institute for Physical Science and Technology \& Joint Center for Quantum Information and Computer Science,  University of Maryland, College Park, MD.}} }
\maketitle

Richard Feynman, who received a Nobel prize in 1965 for his contributions to quantum electrodynamics, famously said that nobody understands quantum mechanics and cautioned against asking: `But how can it be like that?'\footnote{Richard Feynman, \emph{The Character of Physical Law}. The Messenger Lectures, Cornell, 1964. Cambridge: MIT Press, 1965; p. 129.}  Something about the conceptual foundations of the theory is profoundly puzzling, but just what is so disturbing is not easy to pin down---this itself is in dispute. Three books by Olival Freire, Adam Becker, and Philip Ball:

\bigskip

\noindent Olival Freire Jr. \emph{The Quantum Dissidents: Rebuilding the Foundations of Quantum Mechanics (1950--1990),} Foreword by Silvan S. Schweber. xvi + 356 pp., illus., figs., tables, index. Heidelberg: Springer, 2015. \$99.00 (cloth). ISBN 978-3-662-44661-4.
\bigskip

\noindent Adam Becker. \emph{What is Real? The Unfinished Quest for the Meaning of Quantum Physics.}  ix + 370 pp., illus., figs., index. New York: Basic Books, 2018. \$32.00 (cloth). ISBN 978-0-465-09605-3.
\bigskip

\noindent Philip Ball. \emph{Beyond Weird: Why Everything You Thought You Knew About Quantum Physics is Different.} 377 pp, illus., figs., index. London: The Bodley Head, 2018. \pounds 17.99 (cloth). ISBN 978-1-847-92457-5.

\bigskip

\noindent are about the fascinating story of rival attempts to say how it can be like that and, in the case of the Freire and Becker books, about the hostility between opposing camps, which sometimes had devastating consequences for the professional careers of the protagonists. 

The story begins with the publication in 1925 of Werner Heisenberg's breakthrough paper  `On the quantum-theoretical re-interpretation (`Umdeutung') of kinematical and mechanical relations' in \emph{Zeitschrift f\"{u}r Physik}. Shortly afterwards, Heisenberg wrote to Wolfgang Pauli (July 9, 1925):\footnote{Quoted in David C. Cassidy, \emph{Uncertainty: The Life and Science of Werner Heisenberg}. New York: W.H. Freeman, 1992; p. 197.}
\begin{quote}
All of my meagre efforts go toward killing off and suitably replacing the concept of the orbital paths that one cannot observe.
\end{quote}
The orbital paths Heisenberg was referring to were the  `stationary states' of electrons revolving around a central nucleus in Bohr's theory of the atom. Bohr associated electrons jumping between these fixed orbits with  photons emitted or absorbed by an atom, with discrete frequencies corresponding to the observed lines or gaps in atomic spectra.  Heisenberg thought Bohr's orbits, which conflicted with classical electrodynamics as well as classical mechanics, were unphysical. His aim was to get rid of the orbits by replacing classical mechanics with `a theoretical quantum mechanics \ldots in which only relations between observable quantities occur.'\footnote{W. Heisenberg, `Quantum-theoretical re-interpretation of kinematic and mechanical relations.' Translation of `\"{U}ber quantentheoretische Umdeutung kinematischer und mechanischer Beziehungen,' \emph{Zeitschrift f\"{u}r Physik} 33, 879--893 (1925). In B.L. van der Waerden (ed.), \emph{Sources of Quantum Mechanics}.  Amsterdam: North-Holland (1967); p. 262.} 

The popular story is that he accomplished this by the extraordinary manoeuvre of `re-interpreting' classical mechanical quantities like position and momentum as operations represented by arrays, which Max Born identified as matrices when Heisenberg sent him a draft of the paper. The result of applying an operation $A$ followed by an operation $B$ can differ from the result of applying the operations in the reverse order, which is to say that the re-interpreted matrix quantities needn't commute. The actual history of the transition from the commutative physical quantities of classical mechanics to the noncommutative algebra of operators representing `observables' in quantum mechanics is rather more complicated. The basic idea of the `Umdeutung' was motivated by Hans Kramers' dispersion theory, which became part of the 1924 Bohr-Kramers-Slater quantum theory of radiation. Heisenberg extended the procedure in Kramers' dispersion theory to replace position and momentum with corresponding quantum variables, which he took to satisfy the same relations as their classical counterparts---hence the `re-interpretation' of classical quantities.  Bohr's orbits were eliminated by replacing quantities associated with a single orbit with quantities associated with transitions between two orbits, representing the observable features of atomic spectra instead of the unobservable orbits. Multiplying these two-index transition quantities together turned out to be noncomutative: $AB \neq BA$.\footnote{See Michel Janssen, `Arches and scaffolds: bridging continuity and discontinuity in theory change.' In Alan C. Love and William C. Wimsatt (eds.), \emph{Beyond the Meme. Articulating Dynamic Structures in Cultural Evolution}. Minneapolis: University of Minnesota Press, forthcoming 2019; section IV, 4th case study.} 

Heisenberg developed his idea of a noncommutative quantum mechanics in collaboration with  Born and Pascual Jordan, and the first version of quantum mechanics appeared as their \emph{Dreim\"{a}nnerarbeit} paper published later in 1925.  It subsequently became apparent, with Heisenberg's interpretation of his commutation relations in 1927 and the Dirac-Jordan transformation theory, developed independently in 1927 by Jordan and Paul Dirac, that systems in this noncommutative mechanics can't have definite values for all physical quantities simultaneously. In particular, an electron can't have definite position and momentum values and so can't have a well-defined orbit in an atom. 

Erwin Schr\"{o}dinger published a wave-mechanical version of the theory in 1926 that kept the orbits and explained their quantization as a wave phenomenon. It is generally supposed that he also proved the formal equivalence of  wave mechanics and Heisenberg's matrix mechanics, but in fact Schr\"{o}dinger proved only the empirical equivalence of the two theories for experiments relevant at the time. The general theoretical question of equivalence was first addressed  by Jordan and Dirac in their transformation theory and definitively settled by John von Neumann in the first paper of his 1927 trilogy, in which he reformulated quantum mechanics as a theory of  `observables' represented by operators and states represented by rays in Hilbert space.\footnote{Janssen, \emph{ibid.}, section IV, fifth case study. John von Neumann, `Mathematische Begr\"{u}ndung der Quantenmechanik,' K\"{o}nigliche Gesellschaft der Wissenschaften zu G\"{o}ttingen: Mathematisch-physikalische Klasse, Nachrichten 1--57, (1927).} In the second paper of the trilogy, von Neumann derived  the Born rule specifying the probability, in a quantum state, of finding the value of an observable in a given range in a measurement.\footnote{John von Neumann,  `Wahrscheinlichkeitstheoretischer Aufbau der Quantenmechanik,'  K\"{o}nigliche Gesellschaft der Wissenschaften zu G\"{o}ttingen: Mathematisch-physikalische Klasse, Nachrichten 245--272,  (1927).} 

Von Neumann's derivation of the Born rule is usually associated with his proof of the impossibility of `completing' quantum mechanics by  introducing additional `hidden variables'  in his 1932 book.\footnote{John von Neumann, \emph{Mathematische Grundlagen der Quantenmechanik.} Berlin: Springer, 1932.} Since the critique of von Neumann's argument by John Bell in 1965,\footnote{John S. Bell, `On the problem of hidden variables in quantum mechanics,' \emph{Reviews of Modern Physics} 38, 447--452 (1965).} the consensus in the literature seems to be that the proof makes an unwarranted assumption and fails to exclude any significant class of hidden variables, and hence von Neumann's derivation of the Born rule from the geometry of Hilbert space is flawed. I have argued \footnote{Jeffrey Bub, `Von Neumann's ``no hidden variables'' proof: a re-appraisal,' \emph{Foundations of Physics} 40, 1333--1340 (2010).} that Bell's analysis misconstrues von Neumann's `no hidden variables' argument, and that the assumptions are appropriate in the context of the proof. As von Neumann put it, the quantum probabilities are `sui generis'\footnote{John von Neumann, `Quantum logics: strict- and probability-logics.' A 1937 unfinished manuscript published in A.H. Taub (ed.), \emph{Collected Works of John von Neumann}, Volume 4. Oxford and New York: Pergamon Press, 1961;  pp. 195--197.}
 and `uniquely given from the start' as a feature of the geometry of Hilbert space, related to the angle between rays in Hilbert space representing `pure' quantum states.\footnote{John von Neumann, `Unsolved problems in mathematics.' An address to the International Mathematical Congress, Amsterdam, September 2, 1954. In Mikl\'{o}s R\'{e}dei and Michael St\"{o}ltzner (eds.), \emph{John von Neumann and the Foundations of Quantum Physics}. Dordrecht: Kluwer, 2001; pp. 231--245.} 
 
 Physicists, not surprisingly, found wave mechanics more familiar and intuitively appealing than Heisenberg's version of quantum mechanics. The wave theory seemed to provide an explanation of observed phenomena in terms of causal processes evolving continuously in space and time, as opposed to Heisenberg's  derivation of irreducible transition probabilities  from  features of a noncommutative algebraic structure. As Schr\"{o}dinger initially saw it, the wave theory had the advantage of \emph{Anschaulichkeit}, usually translated as clarity or visualizability, here in the specific sense of representability as a causal picture of events unfolding continuously in space and time.\footnote{See B. Cassin, E. Apter, J. Lezra, M. Wood (eds.), \emph{Dictionary of Untranslatables: A Philosophical Lexicon}. Princeton: Princeton University Press, 2014;  p. 37 `Anschaulichkeit,' by Catherine Chavalley.} 

Heisenberg emphatically disagreed. As he put it in a letter to Pauli (June 8, 1926):\footnote{W. Pauli, \emph{Wissentschaftlicher Briefwechsel mit Bohr, Einstein, Heisenberg u.a.}, Volume 1 (1919--1929), A. Hermann, K. von Meyenn, V. F. Weiskopf (eds.).  Berlin: Springer, 1979; p. 328.}
\begin{quote}
The more I think about the physical portion of Schr\"{o}dinger's theory, the more repulsive [abscheulich] I find it. \ldots What Schr\"{o}dinger writes about the Anschaulichkeit of his theory `is probably not quite right,' in other words it's crap [Mist].
\end{quote}
The dispute between Heisenberg and Schr\"{o}dinger about `Anschaulichkeit' is the root of the debate between an orthodox camp, represented primarily by the diverse but related views of Bohr, Heisenberg, Pauli, and Rosenfeld, dubbed the Copenhagen interpretation by Heisenberg, and the dissidents, represented by Schr\"{o}dinger with Einstein as a powerful ally, and later by Bohm, Everett, Bell, and others. How this debate evolved between 1950 and 1990 is the subject of Olival Freire's authoritative and meticulously researched book, \emph{The Quantum Dissidents}.

In a seminal 1935 paper, Einstein, Podolsky, and Rosen drew attention to the existence of non-separable or `entangled' states of separated systems  and exploited their correlations in an argument for the incompleteness of quantum mechanics. Entanglement---what Schr\"{o}dinger called `\emph{the} characteristic trait of quantum mechanics, the one that enforces its entire departure from classical lines of thought'\footnote{E. Schr\"{o}dinger, `Discussion of probability relations between separated systems,' \emph{Mathematical Proceedings of the Cambridge Philosophical Society} 31, 555--563 (1935); p. 555.}--- was pretty much ignored until John Bell came along in the 1960s and re-examined the Einstein-Podolsky-Rosen argument, using  David Bohm's 1951 reformulation in terms of spin components in different directions  rather than position and momentum,\footnote{David Bohm, \emph{Quantum Theory}. New York: Prentice Hall, 1951.} but the question of completeness dominated the debates between Bohr and Einstein. What Einstein had in mind was that something was left out of the quantum theory, which, if added to the theory, would restore the sort of `Anschaulichkeit' characteristic of classical theories.

The completeness issue was never, for Einstein or Schr\"{o}dinger, just about restoring determinism. As Pauli pointed out in correspondence with Born (March 31, 1954):\footnote{M. Born, \emph{The Born-Einstein Letters}. London: Walker and Co., 1971; p. 221. See Christoph Lehner, `Einstein's realism and his critique of quantum mechanics.' In Michel Janssen and Christoph Lehner (eds.), \emph{The Cambridge Companion to Einstein}. Cambridge: Cambridge University Press, 2014.} 
\begin{quote}
Einstein does not consider the concept of `determinism' to be as fundamental as it is frequently held out to be (as he told me emphatically many times) \ldots. In the same way he `disputes' that he uses as criterion for the admissibility of a theory the question: `Is it rigorously deterministic?' Einstein's point of departure is `realistic' rather than `deterministic,' which means that his philosophical prejudice is a different one.'
\end{quote}
The question of  `Anschaulichkeit' morphed into a debate about the possibility of a realist interpretation of quantum mechanics, with the dissidents accusing the Copenhagenists of the sin of positivism or instrumentalism, which by the 1960s had lost much of its appeal among philosophers.

Freire shows how research on the foundations of quantum mechanics, which was not regarded seriously as `real physics'  in the 1950s and 1960s, achieved respectability in spite of sometimes virulent opposition to the dissidents  as the field of quantum information emerged in the 1990s, following Bell's seminal nonlocality result in 1964 and the subsequent tests of quantum nonlocality in the 1980s by Alain Aspect and others. The increasingly sophisticated experimental techniques designed to test aspects of nonlocal entanglement transformed foundational questions from what physicists regarded as the sort of thing one might discuss over a beer at the end of a day of doing real physics, to issues that were taken seriously and could, at least in principle, be settled empirically. 

\emph{The Quantum Dissidents} begins with an account of David Bohm's causal interpretation of quantum mechanics, published as two papers in the \emph{Physical Review} in 1952. I was a graduate student with Bohm's group in London in the 1960s and one of the dissidents mentioned by Freire, so I found the chapter on Bohm particularly fascinating. Influenced by Einstein, Bohm proposed his theory as a counter-example to von Neumann's `no hidden variables' proof, which was widely accepted at the time.  Freire's dscussion illuminates the political and philosophical background to Bohm's theory and the caustic hostility to his theory by Copenhagenists like Leon Rosenfeld, a fellow Marxist. Following his testimony to the House Un-American Activities Committee, where he refused to testify against colleagues, Bohm lost his position at Princeton. He moved to Brazil, where he was offered a position at the University of S\~{a}o Paulo, but was unable to travel after US authorities confiscated his passport. Eventually, after obtaining Brazilian citizenship (and consequently losing his US citizenship), he moved to the Technion in Israel, and finally to the University of London as Chair of Theoretical Physics at Birkbeck College.

The second major dissident in the 1950s was  Hugh Everett III, who proposed a  `relative state' formulation of quantum mechanics, dubbed the `many-worlds' interpretation by Bryce DeWitt. Both Bohm and Everett begin with Schr\"{o}dinger's version of quantum mechanics. For Bohm, the wave function plays the role of a guiding wave for the motions of particles that follow well-defined trajectories. For Everett, the wave function,  as a superposition of terms associated with different possible values of an observable, is all there is as a representation of physical reality. The  superposition  is interpreted as a multiplicity, with each term representing a `branch,' or possible world. As Everett put it to DeWitt (Freire, p. 102): 
\begin{quote}
From the viewpoint of the theory, all elements of a superposition (all `branches') are `actual,' none any more real than another.' It is completely unnecessary to suppose that after an observation somehow one element of the final superposition is selected to be awarded with a mysterious quality called `reality' and the others condemned to oblivion. We can be more charitable and allow the others to exist---they won't cause any trouble anyway because all the separate elements of the superposition (`branches') individually obey the wave equation with complete indifference to the presence or absence (`actuality' or not) of the other elements.
\end{quote}
Probabilities are subjective for an observer in a world (which raises some fundamental issues about probability that remain controversial in contemporary versions of Everett's formulation), and properties, or values of an observable, are relative to a world and not associated with a classical record, as they are on the Copenhagen interpretation. Everett's analysis  accounts for the subjective experience of an observer in this multiverse by considering correlations between memory states.

Everett was a graduate student when he proposed the idea as his PhD dissertation topic. Freire discusses in detail the difficulties Everett faced with his advisor, John Wheeler, who insisted on obtaining Bohr's imprimatur, at one point dispatching Everett to Copenhagen. The visit did not go well.  Years later Rosenfeld wrote to Frederik Belinfante (Freire, p. 114):
\begin{quote}
With regard to Everett neither I nor even Niels Bohr could have any patience with him when he visited us in Copenhagen more than 12 years ago in order to sell the hopelessly wrong ideas he had been encouraged, most unwisely, by Wheeler to develop. He was undescribably (sic) stupid and could not understand the simplest things in quantum mechanics.
\end{quote}

%The last part of the chapter is perhaps the best discussion I have seen of the complete disconnect between Everett's view and the Copenhagenists' understanding of measurement and probability in noncommutative quantum mechanics. Freire's analysis contains much that seems to have been forgotten in contemporary accounts of the measurement problem and the Copenhagen interpretation. His concluding remarks about Everett apply as well to the contemporary debate (Freire, p. 134):
%\begin{quote}
%Everett pointed out what he considered to be the limitations of Bohr's approach and straightforwardly ascribed them to Bohr's dogmatic and conservative stance. There was no effort on his part to reach a deeper understanding of the philosophical background of complementarity, and no hesitation to seek a formulation of quantum mechanics in which Bohr's reflections on the nature of scientific knowledge could be simply bypassed.
%\end{quote}

Another interesting  development was the rival orthodoxy of the Princeton school, represented by Eugene Wigner's treatment of the measurement problem, and the Tausk affair. Von Neumann's Hilbert space formulation of quantum mechanics, accepted as the standard version of the theory, distinguishes two processes: the deterministic Schr\"{o}dinger dynamical evolution that occurs when a system is not measured, and the stochastic `collapse' process formalized by von Neumann's `projection postulate' that occurs during measurement, when the quantum state collapses or is projected onto the term in the superposition representing the measurement outcome. Following von Neumann's analysis, the measurement problem  became the problem of accounting for this dual dynamics, or somehow reconciling the stochastic collapse with the  deterministic Schr\"{o}dinger dynamics. 

Wigner took the position, supported by an argument now known as the `Wigner's friend' argument, that the  collapse of the quantum state is something that happens when information about a measurement outcome enters the mind of a conscious observer. Three Italian physicists, Adriana Daneri, Angelo Loinger, and Giovanni Maria Prosperi (DLP), published a quantum  treatment of macrosystems in 1962 that aimed to show that the collapse is an ergodic process that occurs in the macroscopic measuring instrument. Rosenfeld vigorously defended the DLP theory as vindicating Bohr's view. Klaus Tausk was a young Brazilian physicist at the International Center for Theoretical Physics in Trieste who wrote a paper critical of the DLP theory. Freire shows how Rosenfeld was instrumental not only in suppressing further publication of Tausk's paper, following its appearance as an internal report of the ICTP, but in effectively ruining Tausk's career. After Rosenfeld's intervention, Tausk was abandoned by his advisor and barely obtained his PhD degree. 

I played a small role in the Tausk affair. Tausk sent me a copy of the ICTP report, which formed the basis of my own critique of the DLP theory, published in \emph{Nuovo Cimento} in 1968. According to Freire (Freire, p. 186), I was the only person who cited Tausk's preprint. Since, as Freire documents, I wasn't the only person whom Tausk influenced, the comment shows how shabbily Tausk was treated by the physics community as a direct result of Rosenfeld's intervention.  

%A chapter on the Enrico Fermi summer schools in Varenna in the 1970s, which provided a venue for discussions on the conceptual foundations of quantum mechanics, shows how the changing political climate played a role. 

The final chapters in the book take us up to the 1980s and 1990s, beginning with Aspect's experiments on Bell's theorem and concluding with quantum information as a new and fertile field of investigation in physics, with applications to quantum cryptography, quantum communication, and quantum computation. These developments finally changed the status of research on the conceptual foundations of quantum mechanics from `philosophy' to `real physics.'  Freire quotes Greenberger on the profound significance of Bell's result (Freire, p. 245):
\begin{quote}
John Bell's status in our field has the same [like Isaac Newton, James Watson, and Linus Pauling] mythic quality. Before him there was nothing, only the philosophical disputes between famous old men. He showed that the field contained physics, experimental physics, and nothing has been the same since.
\end{quote}

\emph{What is Real?} by Adam Becker has been phenomenally successful as a non-academic work that covers some of the same ground as Freire. In fact, Becker acknowledges  that Freire's book `easily halved the time I spend on research for this book' (Becker, p. 296). The emphasis, though, is quite different. In \emph{The Quantum Dissidents}, Freire explores how ideas on the conceptual foundations of quantum mechanics evolved in the period after World War II, and the historical, political, and sociological factors that played a role in elevating  quantum foundations from the fringes of physics to a serious research topic. By contrast, Becker's aim is to show how  the `monocracy' of the Copenhagen interpretation was challenged by the dissidents, in spite of great personal loss to their careers as physicists, so that today, while most physicists might still pay lip service to some version of the Copenhagen interpretation, it has become increasingly clear, as Einstein put it, that this was always nothing more than `a tranquilizing philosophy' and `a soft pillow for the true believer' (Becker, p. 14). 

Becker writes vividly in a way that brings the central figures of the story to life: the young Heisenberg and Schr\"{o}dinger inventing quantum mechanics in the 1920s, Einstein and Bohr debating conceptual issues at the Solvay conferences, Heisenberg's arrest in 1945  and internment by the British with other physicists involved in Nazi Germany's nuclear program. The chapters on Bohm and Everett, and the resistance they faced from the Bohr camp, are a great read. The discussion of Bell and his  nonlocality result, and the story about how Clauser, Shimony, Aspect and other figures designed and performed the experiments  in the 1970s and 1980s to test Bell's inequality that were instrumental in bringing about what Aspect calls `the second quantum revolution,' is quite riveting. 

 %Becker acknowledges that Freire's book `easily halved the time I spent on research for this book'  (Becker, p. 296), but Becker's aim is to show how

%The logical positivism of the Vienna Circle appeared on the philosophical scene in the 1920s together with quantum mechanics. Becker argues that the positivist intellectual zeitgeist of the time accounts for the initial broad acceptance of the Copenhagen interpretation as the orthodox interpretation of quantum mechanics, and the emergence and eventual recognition of dissident interpretations became possible with the demise of this philosophical view. 

A core part of Becker's thesis is his dismissal of the Copenhagen interpretation as a positivist or instrumentalist aberration (Becker, pp. 14--15) or, following David Merman (Becker, p. 275), as a retreat to `shut up and calculate' or, quoting David Albert (Becker, p. 283), as simply `incoherent,' `unintelligible,' `gibberish.' In several places Becker invokes the quote, `there is no quantum world,' commonly attributed to Bohr (Becker, p. 14):
\begin{quote}
What does quantum physics tell us about the world? According to the Copenhagen interpretation this question has a very simple answer: quantum mechanics tells us nothing whatsoever about the world. \ldots According to Bohr, there isn't a story about the quantum world because `there is no quantum world. There is only an abstract quantum physical description.'
\end{quote}
The `no quantum world' comment is actually a quote from Bohr's assistant Aage Petersen,\footnote{Aage Petersen, `The Philosophy of Niels Bohr,' \emph{Bulletin of the Atomic Scientists} 19, 8--14 (1963). The quote is on p. 12: `When asked whether the algorithm of quantum mechanics could be considered as somehow mirroring an underlying quantum world, Bohr would answer, ``There is no quantum world. There is only an abstract quantum physical description. It is wrong to think that the task of physics is to find out how nature \emph{is}. Physics concerns what we can \emph{say} about nature.''} who recounts Bohr saying this sort of thing. Bohr probably did make provocative statements along these lines in discussion, but he certainly did not mean that there is simply nothing there, as Becker seems to suggest. 

What could Bohr have meant? Here's my take on it. Quantum mechanics  replaces the commutative algebra of physical quantities of a classical system with a noncommutative algebra of `observables.' This is an extraordinary move, quite unprecedented in the history of physics, and arguably requires us to re-think what counts as an acceptable explanation in physics. To understand  what noncommutativity involves it's helpful to think of two-valued observables. These represent properties of a quantum system (for example, the property  that the energy of the system lies in a certain range of values, with the two values of the observable representing `yes' or `no'), or propositions (the proposition asserting that the value of the energy lies in a certain range, with the two values representing `true' or `false'). The two-valued observables of a classical system form a Boolean algebra. 

We all, in a loose sense, understand the concept of a Boolean algebra, even if the term is unfamiliar. It's simply a formalization of the way in which we commonly think of properties or propositions fitting together when we combine them with connectives `and,' `or,' `not,' so that it's possible to imagine a classical or commonsense `state of reality'  in which every proposition is assigned a truth value, either `true' or `false,' consistently with the connectives. George Boole characterized the algebraic structure he identified in  1847  as capturing `the conditions of possible experience.'\footnote{Itamar Pitowsky. `George Boole's ``conditions of possible experience'' and the quantum puzzle.' \emph{British Journal for the Philosophy of Science} 45, 9--125 (1994).}

To say that the algebra of observables of a quantum system is noncommutative is formally equivalent to saying that  the sub-algebra of properties or propositions is non-Boolean. Hilbert space formalizes this non-Booleanity in a particular way.  The Boolean algebra of classical mechanics is replaced by a collection of  Boolean algebras,  one for each set of commuting two-valued observables. The interconnections of commuting and noncommuting observables preclude the possibility of embedding the whole collection into one inclusive Boolean algebra, so you can't assign truth-values consistently to the propositions about observable values in all these Boolean algebras. Putting it differently, there are Boolean algebras in the collection of Boolean algebras of a quantum system, for example the Boolean algebras for position and momentum, or for spin components in different directions, that don't fit together into a single Boolean algebra, unlike the corresponding collection for a classical system. A `world' in which it is possible to talk about truth and falsity, referring to \emph{this} rather than \emph{that}, is Boolean.  In this sense, there is no non-Boolean quantum `world'---there is no consistent assignment of truth-values to all quantum propositions defining a `state of reality.' 

Bohr did not refer to Boolean algebras, but the concept is simply a precise way of codifying a significant aspect of what Bohr meant by  his constant  insistence that `the account of all evidence must be expressed in classical terms,' that's to say `unambiguous language with suitable application of the terminology of classical physics,' for the simple reason, as he put it, that we need to be able `to tell others what we have done and what we have learned.'\footnote{Niels Bohr, `Discussions with Einstein on epistemological problems in modern physics.' In P. A. Schilpp (ed.), \emph{Albert Einstein: Philosopher-Scientist}, The Library of Living Philosophers, Volume 7 (Open Court, Evanston, 1949); pp. 201--241.} Formally speaking, the significance of `classical' here as being able `to tell others what we have done and what we have learned' is that the events in question should fit together as a Boolean algebra, so conforming to Boole's `conditions of possible experience.'

In the non-Boolean theory of quantum mechanics, probabilities arise via  the Born rule as a feature of the geometry of Hilbert space. These probabilities can't be understood as quantifying ignorance about the pre-measurement value of an observable, as in a Boolean theory. For the Copenhagenists, quantum mechanics is a new sort of non-representational theory for an irreducibly indeterministic universe, with a new type of nonlocal probabilistic correlation  for `entangled' quantum states  of separated systems, where the correlated events are  intrinsically random, not merely apparently random like coin tosses. 

Hilbert space provides the kinematic framework for such an irreducibly indeterministic universe, in a similar sense to which Minkowski space provides the kinematic framework for the physics of  a non-Newtonian, relativistic universe. In special relativity, Lorentz contraction is  a kinematic effect of the spatio-temporal constraints on events imposed by the geometry of Minkowski space. In quantum mechanics, the loss of information in a quantum measurement---Bohr's `irreducible and uncontrollable disturbance'---is  a kinematic (i.e., pre-dynamic)  effect of \emph{any} process of gaining information of the relevant sort, irrespective of the dynamical processes involved in the measurement process,  given the objective probabilistic constraints on correlations between events imposed by the  geometry of Hilbert space.

On this view, quantum mechanics does not provide a representational  explanation of events. Noncomutativity or non-Booleanity makes quantum mechanics quite unlike any theory we have dealt with before, and  there is no reason, apart from tradition, to assume that a theory should provide   the sort of `anschaulich'  explanation we are familiar with in a  theory that is commutative or  Boolean at the fundamental level.

But aren't Bohm's theory and the Everett interpretation  perfectly good representational or `anschaulich' theories, empirically equivalent to standard quantum mechanics? After all, they are proposals intended to explain phenomena in terms of a causal process in space and time involving an underlying ontology. It's questionable, though, whether they are in fact good representational theories, or even whether they make the same predictions as standard quantum mechanics in all experimental setups: the devil is in the details. 

As Bell points out,\footnote{J. S. Bell, \emph{Speakable and Unspeakable in Quantum Mechanics}. Cambridge: Cambridge University Press, 1987; p. 11.} Bohm's theory involves action at a distance at the level of the hidden variables:  `an explicit causal mechanism exists whereby the disposition of one piece of apparatus affects the results obtained with a distant piece,' so that `the Einstein-Podolsky-Rosen paradox is resolved in the way which Einstein would have liked least.' The problem of making sense of probability in an Everettian universe, where everything that can happen does happen in some world, is still a contentious issue. Everettian quantum mechanics and Bohm's theory in a generalized sense can be understood as different interpretations of the same underlying theory---Everett's theory is simply `the pilot-wave  theory without trajectories,' as Bell put it\footnote{J.S.Bell, \emph{ibid.}, p. 133.}--- and there is a sense in which this Bohm-Everett theory is not empirically equivalent to standard quantum mechanics. For scenarios that involve `Wigner's friend' type setups, these theories make different predictions, as pointed out recently by Veronika Baumann and Stefan Wolf.\footnote{Veronika Baumann and Stefan Wolf, `On formalisms and interpretations,' \emph{Quantum} 2, 99 (2018)} Such experiments are not remotely possible in practice, although nothing in quantum mechanics forbids their implementation in principle. So perhaps the Copenhagenists' riposte to Mermin's characterization of the Copenhagen interpretation as `shut up and calculate' should be `shut up and wait' until physicists figure out how to do a crucial experiment that decides between these theories.

%The Freire and Becker books are both contributions to the history of physics. In spite of my caveats about Becker's treatment of the Copenhagen interpretation, \emph{What is Real?} is well worth reading, not only because it's a grippingly told story, but also for the questions Becker raises about what constitutes a good explanation in physics.  

%There is no list of chapters in \emph{Beyond Weird}. This makes it a little difficult to navigate the book, which reads like an extended essay, broken into short excerpts marked off by 2-page spreads with a white title on a grey background. Perhaps this was Ball's intention, to keep the reader fully immersed in the conceptual problems and the dialectic of their resolution. 

Philip Ball's book, \emph{Beyond Weird}, is a non-technical but nonetheless  clear and deeply illuminating exploration of what makes quantum mechanics so tantalizingly puzzling. The discussion of various interpretations includes Bohm's theory and Everett's many-worlds interpretation, but also QBism (short for quantum Bayesianism), an information-theoretic interpretation developed primarily  by Chris Fuchs and R\"{u}diger Schack, based on treating quantum probabilities as purely subjective degrees of belief (pp. 121--125). (Surprisingly, there is no reference to QBism in Becker's book, although there is a cryptic reference to Fuchs in the Acknowledgements.) 

Ball is not a fan of the many-worlds interpretation (Ball, p. 305):
\begin{quote}
What quantum theory seems to insist is that at the fundamental level the world cannot supply clear `yes/no' empirical answers to all the questions that seem at face value as though they should have one. The calm acceptance of that fact by the Copenhagen Interpretation seems to some, and with good reason, to be far too unsatisfactory and complacent. The MWI [many-worlds interpretation] is an exuberant attempt to rescue the `yes/no,' albeit at the cost of admitting both of them at once. \ldots  And in the end, if you say everything is true, you have said nothing.
\end{quote}

In a perceptive  analysis of the Einstein-Poldolsky-Rosen argument and Bell's  nonlocality proof, Ball draws the opposite conclusion to Becker about the Copenhagen interpretation (Ball, p. 177):
\begin{quote}
There can be no clearer demonstration that Bohr's refusal to contemplate any meaning for things not observed wasn't just stubborn pedantry. If he's right, there are \emph{measurable consequences}. This wouldn't mean that the Copenhagen interpretation is correct, but it would mean that Einstein's hidden variables---any attempt, in fact, to fix everything in principle in a quantum system before it is observed---won't wash.
\end{quote}
He goes on to make an insightful distinction between action at a distance and quantum nonlocality (Ball, p. 184):
\begin{quote}
But this nonlocality is just what quantum entanglement undermines---which is why `spooky action at a distance is precisely the wrong way to look at it. \ldots In quantum mechanics, properties can be \emph{non-local}. Only if we accept Einstein's assumptions of locality do we need to tell the story in terms of a measurement on particle $A$ `influencing' the spin of particle $B$. Quantum non-locality is the \emph{alternative} to that view.
\end{quote}

There is a lengthy and sophisticated discussion of  how decoherence plays a role in ensuring the effective classicality of  the macrolevel (Ball, pp. 198--251). The final chapters contain an excellent non-technical discussion of  information-theoretic approaches to quantum mechanics, and an explanation of how entanglement can be exploited for efficient quantum communication and quantum computation (Ball, p, 284):
\begin{quote}
Understanding exactly how quantum mechanics can improve computing may turn out to provide insights into one of the deepest questions of the field: what quantum information really is, and how it can be transmitted and altered.
\end{quote}

So read all three books. In different ways, they are all about the ongoing saga of interpreting the quantum world, and about  how we are beginning to get some fascinating new answers to old questions.

\end{document}